\documentclass[prl,twocolumn]{revtex4}
\usepackage{amsmath,amssymb}
\usepackage{epsfig}
\usepackage{color}
\usepackage{fixltx2e}
\usepackage[normalem]{ulem}

\newcommand {\ra}{\rightarrow}

\newcommand{\bsf}{\begin{subfigure}} 
\newcommand{\esf}{\end{subfigure}} 

\graphicspath{{eps/}}            

\begin{document}
\title{
Bi-solitons on the surface of a deep fluid: an inverse scattering transform perspective based on perturbation theory}

\author{Andrey Gelash$^{1}$}\email{Andrey.Gelash@u-bourgogne.fr}
\author{Sergey Dremov$^{3}$}
\author{Rustam Mullyadzhanov$^{2,4}$}
\author{Dmitry Kachulin$^{2,3}$}

\affiliation{$^{1}$Laboratoire Interdisciplinaire Carnot de Bourgogne (ICB), UMR 6303 CNRS -- Université Bourgogne Franche-Comté, 21078 Dijon, France}
\affiliation{$^{2}$Novosibirsk State University, Novosibirsk 630090, Russia}
\affiliation{$^{3}$Skolkovo Institute of Science and Technology, Moscow 121205, Russia}
\affiliation{$^{4}$Institute of Thermophysics SB RAS, Novosibirsk 630090, Russia}

\begin{abstract}
We investigate theoretically and numerically the dynamics of long-living oscillating coherent structures -- bi-solitons -- in the exact and approximate models for waves on the free surface of deep water.
We generate numerically the bi-solitons of the approximate Dyachenko–Zakharov equation and fully nonlinear equations propagating without significant loss of energy for hundreds of the structure oscillation periods, which is hundreds of thousands of characteristic periods of the surface waves.
To elucidate the long-living bi-soliton complex nature we apply an analytical-numerical approach based on the perturbation theory and the inverse scattering transform (IST) for the one-dimensional focusing nonlinear Schrödinger equation model.
We observe a periodic energy and momentum exchange between solitons and continuous spectrum radiation resulting in repetitive oscillations of the coherent structure.
We find that soliton eigenvalues oscillate on stable trajectories experiencing a slight drift on a scale of hundreds of the structure oscillation periods so that the eigenvalue dynamic is in good agreement with predictions of the IST perturbation theory.
Based on the obtained results, we conclude that the IST perturbation theory justifies the existence of the long-living bi-solitons on the surface of deep water which emerge as a result of a balance between their dominant solitonic part and a portion of continuous spectrum radiation.
\end{abstract}
\maketitle
%
%

%
%
Formation of stable localized coherent structures -- solitons -- is one of the key evolution scenarios of nonlinear wave systems \cite{zakharov2012solitons}.
When such a system is Hamiltonian, solitons emerge due to a balance between nonlinearity and dispersion, while in non-conservative cases, an additional balance between energy gain and loss comes into play \cite{zakharov2012solitons,remoissenet2013waves,ankiewicz2008dissipative}.
Being described by nonlinear partial differential equations (PDEs), systems with solitons can be seen in almost all fields of physics, for example, in hydrodynamics, optics, and plasmas \cite{kivshar2003optical,remoissenet2013waves}.
While individual stationary solitons are ubiquitous for nonlinear wave models, long-living multi-soliton complexes are not so common and thus draw particular attention and are of great interest for experimental implementation.
For example, a bound state of solitons has been observed in mode-locked fiber lasers, Bose-Einstein condensates, and specially designed optical waveguides \cite{soto2004bifurcations,al2011formation,stratmann2005experimental,grelu2012dissipative}.
For a Hamiltonian wave model the presence of recursive multi-soliton behavior might be a signature of its integrability or nearly integrable dynamics \cite{NovikovBook1984,KivsharRMP1989,faddeev2007hamiltonian,berman2005fermi}. 
The inverse scattering transform (IST) theory elucidates the particle-like features of solitons in exactly integrable nonlinear PDEs by proving that solitons correspond to the time-invariant eigenvalue spectrum of an auxiliary scattering problem \cite{NovikovBook1984,AblowitzBook1981}.
For example, solitons of the integrable one-dimensional nonlinear Schr\"odinger equation (NLSE) collide elastically forming bouncing multi-soliton complexes and preserve their parameters during the whole system evolution  \cite{zakharov1972exact}.
When integrability is broken by adding weak extra terms to the model, solitons can still form a long-living, but usually inelastic complexes, which dynamics is described by the IST perturbation theory \cite{buryak1994internal,besley2000soliton,KivsharRMP1989}.

We consider the Hamiltonian models of the 2D hydrodynamics with a free surface: (i) focusing NLSE \cite{zakharov1968stability}, (ii) Dyachenko-Zakharov envelope equation (DZE) \cite{dyachenko2017envelope}, and (iii) fully nonlinear equations for the $R$-$V$ variables (RVE) \cite{ovsyannikov1973dynamika,tanveer1991singularities,dyachenko1996analytical, dyachenko2001dynamics}. These models are the members of the Hamiltonian hierarchy of equations for the free surface water waves \cite{zakharov1968stability,dyachenko2017super}, in which the NLSE describes only the weakly nonlinear narrow-banded wave trains while the DZE captures many of the nonlinear effects presented in the full model \cite{zakharov1968stability,dyachenko2017envelope}. Comparative analysis of the behavior of wave groups in the approximate DZE and the exact RVE models provides insights into how the model objects are expected to be seen in nature \cite{slunyaev2013highest,onorato2013rogue,tikan2022prediction}. The original equations describing 2D hydrodynamics of deep water waves propagating on the free surface of an ideal incompressible fluid in the presence of gravity represent the Laplace equation with kinematic and dynamic boundary conditions at the surface:
\begin{eqnarray}\label{FreeSurfEq}
&&\phi_{xx} + \phi_{yy} = 0,  \,\quad\text{with}\,\, \phi_y \ra 0, \,\, \text{at} \,\, y \ra -\infty,  \cr
&&\eta_t + \eta_x \phi_x  = \phi_y, \,\,\, ~\text{at}~ y=\eta,  \cr
&&\phi_t + (\phi^2_x + \phi^2_y) / 2 +g\eta  = 0, \,\,\,\,\,\, ~\text{at}~ y=\eta \,.
\end{eqnarray}
Here $x$ and $y$ are horizontal and vertical coordinates, $t$ is time, $g$ is the free-fall acceleration,
$\eta(x,t)$ is the shape of the surface, $\phi(x,y,t)$ is a hydrodynamic potential inside the fluid. Classical problem (\ref{FreeSurfEq}) has been known since the 19th century \cite{lamb1924hydrodynamics} and nowadays represent a backbone of theoretical, numerical, and experimental studies \cite{zakharov1992kolmogorov,pelinovsky2008book,osborne2010book,ducrozet2016hos}.

The Hamiltonian of system (\ref{FreeSurfEq}) is \cite{zakharov1968stability}:
\begin{eqnarray}\label{Hamiltonian:total}
H &=& \frac{1}{2} \int dx \int_{-\infty}^\eta |\nabla\phi|^2 dy +  \frac{g}{2}\int \eta^2 dx.
\end{eqnarray}

For numerical solving of the fully nonlinear equations we apply conformal mapping of the fluid domain $z=x+iy$ confined by a free boundary onto the lower half-plane of the new complex variable $w=u+iv$ at $v\leq 0$. In terms of special analytical functions $R(u,t)$ and $V(u,t)$ original equations (\ref{FreeSurfEq}) turn into the RVE:
\begin{eqnarray}\label{RVEq}
&&R_t = i (U R_w - R U_w), \cr
&&V_t = i (U V_w - R B_w) +g (R-1),
\end{eqnarray}
with boundary conditions: $R \ra 1$, $V \ra 0$ at $v \ra -\infty$; see \cite{ovsyannikov1973dynamika,tanveer1991singularities,dyachenko2001dynamics} for the conformal mapping technique and Eq.~(\ref{RVEq}) derivation. We define here $U = \widehat{P} ( V R^* + V^*R )$ and $B =  \widehat{P}(V V^*)$ where $\widehat{P} = ( 1 + i \widehat{H}) / 2$, and $\widehat{H}$ is the Hilbert transform.

Assuming the wave steepness to be small, $\mu = \eta_x \ll 1$, one can expand Hamiltonian (\ref{Hamiltonian:total}) up to the fourth order of $\eta(x,t)$ and $\phi(x,\eta,t)$, and find the DZE for an approximate description of the water wave train in terms of canonical complex envelope variable $C(x,t)$ \cite{dyachenko2017envelope},
\begin{eqnarray}\label{EnvEq}
i\frac{\partial C}{\partial t} &=& \hat\theta_{k_0+k}\Bigg[\left(\omega_{k_0+k}-\omega_{k_0}-
\frac{\partial \omega_{k_0}}{\partial k_0} \hat k \right) C 
\\\nonumber
&+& k_0^2\left(|C|^2C\right) + i\frac{\partial}{\partial x}\left(\hat k(|C|^2)C +i|C|^2\frac{\partial C}{\partial x}\right) 
\\\nonumber
&-& i k_0\left(C\frac{\partial }{\partial x}|C|^2 
+2|C|^2\frac{\partial C}{\partial x} -i \hat k(|C|^2)C\right)
\Bigg].
\end{eqnarray}
Here, operator $\hat k$ is multiplication by $|k|$ in Fourier space, $\hat \theta_k$ is the Heaviside step function, $k_0$ is an arbitrary characteristic wavenumber and $\omega_{k_0}=\sqrt{g k_0}$ is the corresponding linear frequency related to characteristic period of the waves $\tau_0 = 2 \pi/\omega_{k_0}$. Models (\ref{RVEq}) and (\ref{EnvEq}) being advantageous for analytical and numerical treatment are used in fundamental studies and find applications in deterministic wave forecasting \cite{fedele2014certain,dyachenko2021short,dyachenko2022free,stuhlmeier2021deterministic}.

Under the additional assumption of narrow band wave spectrum, we obtain the remaining model of our hierarchy -- the NLSE written in terms of complex envelope variable $q(x,t)$,
\begin{eqnarray}
\label{eqNLSzero}
i q_t + q_{xx} / 2 + |q|^2 q = 0.
\end{eqnarray}

Being integrable, the NLSE exhibits exact multi-soliton solutions, $q^{\text{S}}_{N}(x,t)$, where $N$ is the number of solitons. The simplest single soliton solution represents the well-known expression,
\begin{eqnarray}
	q_{1}^{\mathrm{S}}(x,t) = a_{1} \frac{\exp\big[iv_{1}(x-x_{1}) + i(a_{1}^{2}-v_{1}^{2})t/2 + i\theta_{1}\big]}{\cosh a_{1} \big(x-v_{1}t-x_1\big)},
	\label{1-SS}
\end{eqnarray}
with real-valued parameters $a_{1}$, $v_{1}$ for soliton amplitude and velocity and $\theta_{1}$ and $x_1$ for its phase and position.

Meanwhile, numerical works revealed solitary waves for the DZE and RVE models \cite{dyachenko2008formation,slunyaev2009numerical} observed later in water wave tank experiments \cite{slunyaev2013simulations,slunyaev2017laboratory}.
 In addition, recent numerical studies discovered extremely long-living bi-solitons in both DZE and RVE models, which oscillate without significant loss of energy hundreds of the structure periods $T$, which is $\sim 10^{5}\,\tau_0$ in dimensional units \cite{kachulin2020multiple,kachulin2021bound}.
The theoretical description of such recursive coherent complexes on the surface of a deep fluid, their internal structure, and interaction mechanisms remains an open question.
To understand the behavior of the long-living bi-solitons in the DZE and RVE, we propose an analytical-numerical approach based on the IST theory for our theoretical benchmark model -- the NLSE.

\begin{figure}[t]\centering
	\includegraphics[width=1.0\linewidth]{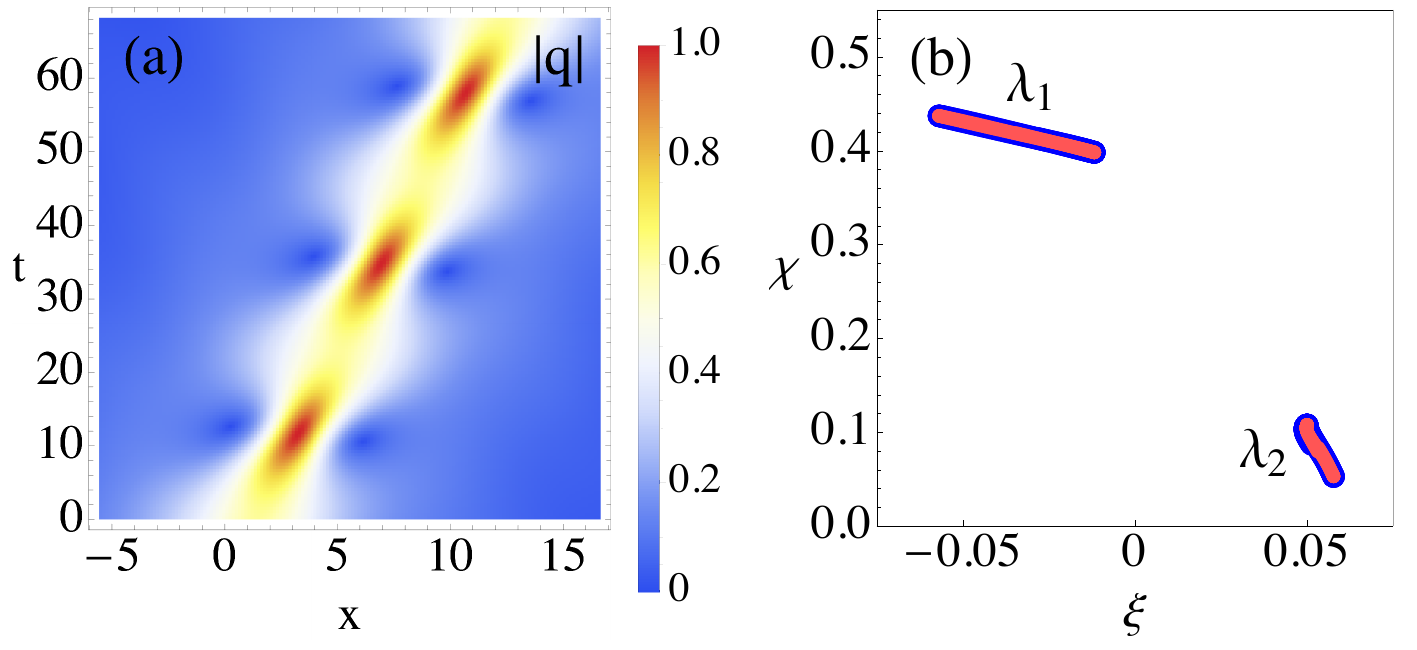}
	\caption{\small Nonlinear behavior of the DZE bi-soliton. (a) Spatio-temporal diagram of the wavefield envelope $|q_{\text{(DZE)}}|$ with characteristic oscillation period $T \approx 24.5$.  (b) Stable trajectories of soliton eigenvalues obtained as a union of the computed time series of $\lambda(t_j)$. Blue and red lines correspond to two complete cycles of the bi-soliton oscillations separated by $200 T$.
	}
	\label{fig:fig1}
\end{figure}

We generate long-living bi-solitons of the DZE and RVE numerically similar to that in \cite{dyachenko2008formation,slunyaev2009numerical,kachulin2021bound}.
We set initial conditions using the two-soliton solution $q^{\text{S}}_{2}(x)$ of the NLSE characterized by different soliton amplitudes $a_1 \ne a_2$ and zero velocities $v_{1,2} = 0$. Then, we substitute such standing bi-soliton complex into the considered equations. Not being a solution to the DZE and RVE models, the initial structure emits incoherent waves at the beginning of the evolution. We absorb these waves by damping at the edges of the computational domain, allowing the structure to find its stable, long-living state.
Later, we turn off the damping and observe that the remaining structure propagates stably for hundreds of characteristic structure periods without losing energy.
We present an example of spatiotemporal oscillating dynamics of the DZE bi-soliton having $T \approx 24.5$ in Figure \ref{fig:fig1}(a) in terms of dimensionless wavefield envelope.
In addition, Figure \ref{fig:fig4} shows the free surface profiles in dimensional units of bi-soliton in the RVE at minimum and maximum amplitude.
More examples of the bi-soliton dynamics including animation of the wavefield evolution as well as details on the numerical procedure used are given in Supplemental Material \cite{SuppM}.

Our approach to analyzing bi-solitons starts with writing the NLSE for the complex-valued wavefield envelope $q(x,t)$ with the right-hand side (RHS) in a general form:
\begin{eqnarray}
\label{eqNLS}
\text{NLSE}[q(x,t)]  &=&  \text{RHS}[q(x,t)],
\\\nonumber
\text{NLSE}[q(x,t)] &\equiv& i q_t + q_{xx} / 2 + |q|^2 q.
\end{eqnarray}

We obtain the DZE/RVE wavefields in terms of $q(x,t)$ by applying the corresponding transformations between models (\ref{RVEq}), (\ref{EnvEq}) and the NLSE,
\begin{equation}
\label{q_transformation}
    q_{\text{(DZE)}} = q\{C(x,t)\}, \quad  q_{\text{(RVE)}} = q\{R(u,t),V(u,t)\}.
\end{equation}
All details of transformations (\ref{q_transformation}) can be found in Supplementary Materials \cite{SuppM}. Now we can directly substitute $q_{\text{(DZE/RVE)}}$ into the NLSE and calculate not zero, but residual which is exactly the RHS for (\ref{eqNLS}):
\begin{eqnarray}
\label{RHS_forms}
\text{RHS}_{\text{(DZE/RVE)}} = \text{NLSE}[q_{\text{(DZE/RVE)}}].
\end{eqnarray}

When $\text{RHS} \equiv 0$, system (\ref{eqNLS}) is integrable and associated with the following linear auxiliary Zakharov--Shabat system for a vector wave function $\Phi = (\phi_1,\phi_2)^{\text{T}}$:
\begin{eqnarray}
\widehat{\mathcal{L}} \Phi = \lambda \Phi, \quad\quad\quad \widehat{\mathcal{L}} = 
\begin{pmatrix}
i \partial_x   & -i q(x) \\ -i q^*(x)   & - i \partial_x
\end{pmatrix},
\label{ZSsystem}
\end{eqnarray}
where $\lambda = \xi + i \chi$ is the time-independent complex spectral parameter, while the superscripts $\text{T}$ and the star stand for a transposition and complex conjugation.
As typically done, we consider Eq. (\ref{ZSsystem}) at a fixed moment of time $t_0$ with $q(x) = q(t_0, x)$ playing the role of a potential. Solving the scattering problem for the system (\ref{ZSsystem}), one finds the wavefield IST spectrum (scattering data) consisting of a set of discrete eigenvalues and norming constants $\{\lambda_n,\,\rho_n\}$, $n = 1,..., N$ with $\chi_n>0$ and the reflection coefficient $r(\xi)$. The first -- discrete part of the IST spectrum corresponds to $N$ solitons having parameters connected to the set $\{\lambda_n,\,\rho_n\}$ as,
\begin{eqnarray}
\label{C_param_S}
    \xi_n &=& v_n/2, \quad \chi_n = a_n/2,
    \\\nonumber
	\rho_{n} &=& 2i\chi_n\exp\big[i\pi - 2i\lambda_{n}x_{n} - i\theta_{n}\big].
\end{eqnarray}
Meanwhile the reflection coefficient $r(\xi)$ being associated with the continuous part of the operator $\widehat{L}$ spectrum describes nonlinear dispersive radiation.

The IST theory proves that $\{\lambda_n\}$ and $|r(\xi)|$ do not change when the wave field $q(x)$ evolves according to the NLSE and only soliton phases and positions, and the phases of the radiation change trivially in time \cite{NovikovBook1984,faddeev2007hamiltonian}. As such, the IST spectrum represents a nonlinear analog of conventional Fourier harmonics. It is used as a powerful tool in analyzing nonlinear wave fields, including the water surface ones \cite{OsborneBook2010, randoux2018nonlinear, chekhovskoy2019nonlinear, suret2020nonlinear, turitsyn2020nonlinear, slunyaev2021persistence, teutsch2022contribution, tikan2022prediction, agafontsev2023bound}. When $r(\xi)=0$ the wave filed is composed of solitons only and its evolution can be described with the exact $N$-soliton solution $q^{S}_{N}(x,t)$; see \cite{NovikovBook1984,matveev1991darboux} and Supplemental Material \cite{SuppM} for background on the IST formalizm and exact multi-soliton formulas.

\begin{figure}[!t]\centering
	\includegraphics[width=0.99\linewidth]{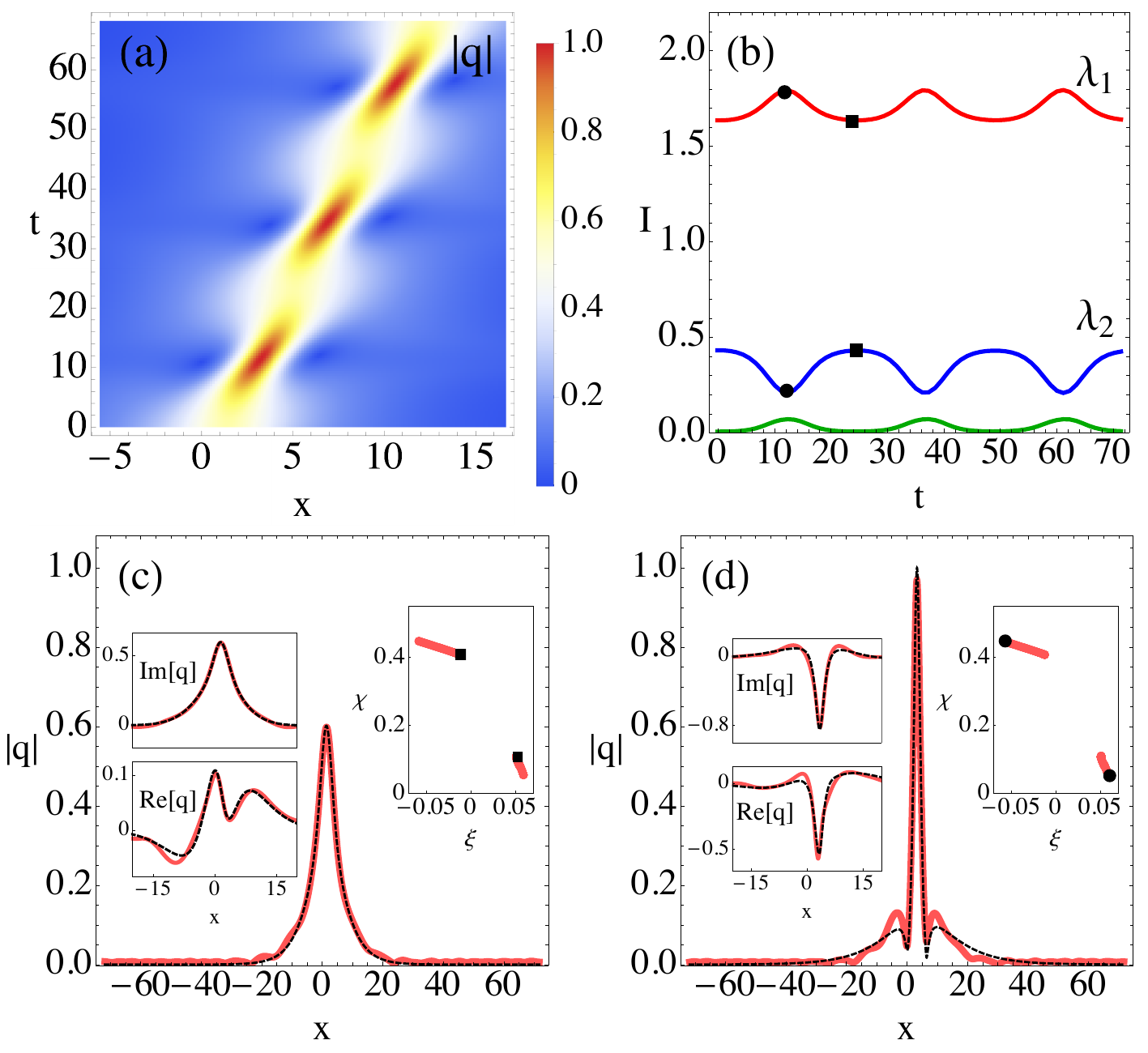}
	\caption{\small Further IST analysis of the DZE bi-soliton. (a) Spatiotemporal dynamics of the two-soliton model (\ref{2ss_model}). (b) Time-evolution of the integral of motion $I$ for each component of the IST spectrum: red and blue lines correspond to $\lambda_1$ and $\lambda_2$ respectively, while the green line shows the impact of continuous spectrum radiation. Panels (c) and (d) show the DZE wavefield (red solid lines) at minimum (c) and maximum (d) amplitude, together with the NLSE two-soliton model wavefield shown by black dashed lines. Insets in (c) and (d) show real and imaginary parts of the wavefields and also indicate the eigenvalue locations by black squares and dots.
	}
	\label{fig:fig2}
\end{figure}

In a general case when $\text{RHS} \ne 0$, system (\ref{eqNLS}) is no longer integrable and the IST eigenvalues are not stationary.
However, when the NLSE part in (\ref{eqNLS}) dominates on the $\text{RHS}$, one can apply the perturbation theory and express the evolution of the eigenvalues \cite{KaupSIAM1976, KarpmanJETP1977}:
\begin{eqnarray}
\frac{\partial \lambda}{\partial t} = \frac{\langle \Phi^\dagger, \widehat{\text{RHS}}~\Phi \rangle}{\langle \Phi^\dagger, \Phi \rangle},
\quad
\widehat{\text{RHS}} = 
\begin{pmatrix}
0     &    -i \text{RHS}    \\     i \text{RHS}    &     0
\end{pmatrix},
\label{dzeta}
\end{eqnarray}
where $\Phi^\dagger = (\phi^*_2,\phi^*_1)^{\text{T}}$ and the scalar product $\langle f, g \rangle = \int_{-\infty}^{\infty} f^* g dx$. Formulation (\ref{eqNLS}) together with equations (\ref{ZSsystem}) and (\ref{dzeta}) are the basis of the classical IST perturbation theory for which many exactly solvable cases of certain RHSs have been studied previously; see \cite{KivsharRMP1989,YangBook} and also some recent works \cite{coppini2020effect,mullyadzhanov2021solitons}. However, many physically important, nearly integrable systems are left without consideration because their RHS is too complicated. In our approach we do not require the $\text{RHS}$ term in an explicit form and instead evaluate it numerically using Eq.~(\ref{RHS_forms}).

We deal with discrete wavefields $q_{\text{(DZE/RVE)}}(x_i,t_j)$ obtained from simulations of the DZE and RVE and use a combination of analytical and numerical IST tools to analyze them.
We begin with bi-solitons in the DZE and, at first, solve the scattering problem for a series of time-evolving wavefield profiles numerically using standard algorithms supplemented by our recent developments \cite{BofOsb1992, Mullyadzhanov2019, Gelash2020}; see details in Supplemental Material \cite{SuppM}. We find two discrete eigenvalues $\lambda_1(t_j)$ and $\lambda_2(t_j)$ and non-zero reflection coefficient $r(\xi,t_j)$ as functions of discrete time steps $t_j$, with $\Delta t = t_{j}-t_{j-1} \approx 0.1$. The eigenvalues corresponding to solitons of the bi-soliton structure oscillate on stable trajectories during hundreds of $T$; see Figure~\ref{fig:fig1}(b). Note, that solitons have nonzero velocities describing by the real part of $\lambda$. With the computed full set of scattering data, we represent the wavefield at each moment of time as two NLSE solitons and continuous spectrum radiation. To measure the impact of each scattering data component in the wavefield composition and energy, we use the NLSE integral of motion,
\begin{equation}
\label{Nintegral}
    I = \int_{-\infty}^{\infty} |q|^2 dx ,
\end{equation}
which can be evaluated individually for discrete spectrum eigenvalues as $I^{\text{DS}}$ and continuous spectrum $I^{\text{CS}}$ within the IST theory \cite{NovikovBook1984}, so that $I = I^{\text{DS}} + I^{\text{CS}}$; see details in Supplementary Materials \cite{SuppM}.

\begin{figure}[!t]\centering
	\includegraphics[width=1.0\linewidth]{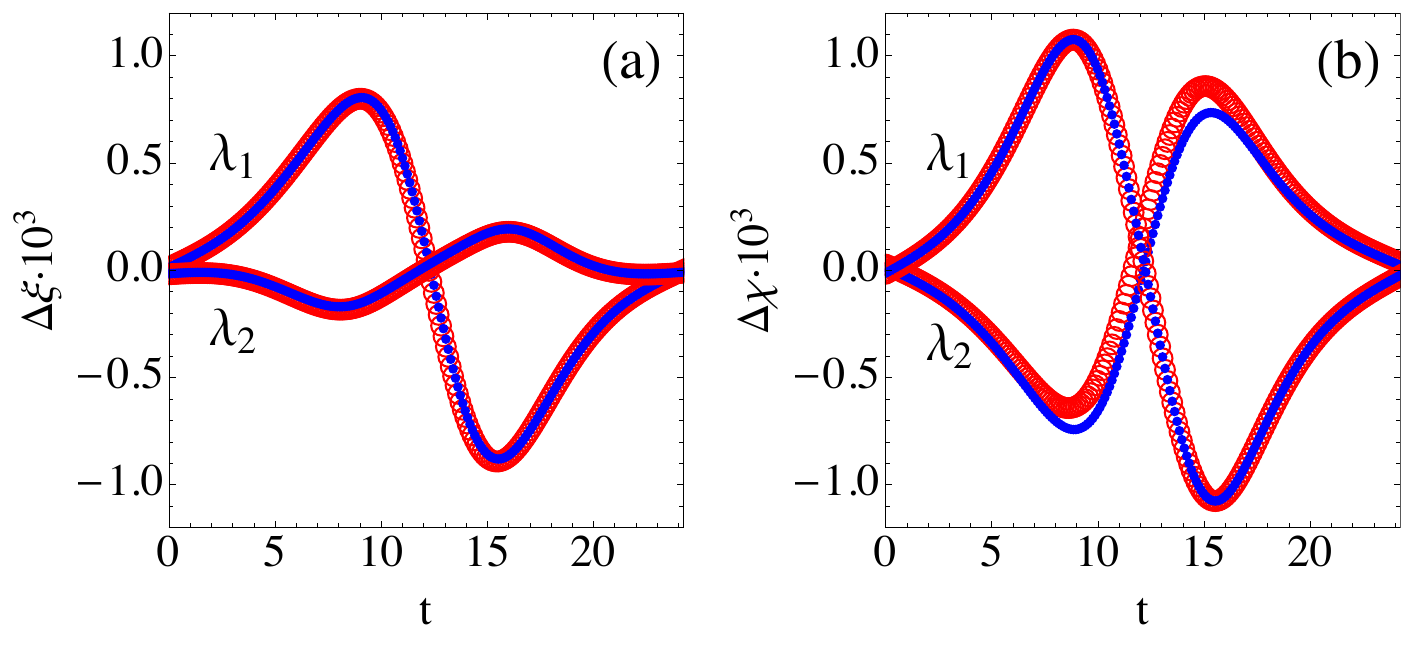}
	\caption{\small Changes of discrete eigenvalues $\Delta\lambda_j = \Delta \xi_j + i \Delta \chi_j$ at each time step $t_j$ within one oscillation period $T$ of the DZE bi-soliton. Panels (a) and (b) correspond to $\Delta \xi$ and $\Delta \chi$ respectively. Blue dots show the direct numerical computation of soliton eigenvalues changes for each subsequent wavefield $q(t_j)$ and $q(t_{j+1})$. Red circles correspond to predictions of the perturbation theory provided by Eq.~(\ref{dzeta}).
	}
	\label{fig:fig3}
\end{figure}

At the edge trajectory points, the bi-soliton exhibits minimum or maximum of its intensity; see Figure~\ref{fig:fig2}. We find that at the minimum intensity configuration, the impact of the continuous spectrum is negligible so that $I^{\text{CS}}/I \approx 0.005$, and the wave field represents almost the exact NLSE bi-soliton; see Figure~\ref{fig:fig2}(a). 
During the structure evolution, the role of the radiation increases, reaching its maximum $I^{\text{CS}}/I \approx 0.036$ at the other edge point; see Figure~\ref{fig:fig2}(b).
In other words an NLSE bi-soliton taken at the minimum intensity configuration evolves as a long-living oscillating complex stabilized by a minor radiation gradually increasing up to the high-amplitude wavefield configuration.
The two solitons and radiation are in periodic energy exchange, which we demonstrate in Figure~\ref{fig:fig2}(b) using time dependence of $I$ for each part of the scattering data. 
We measure the radiation only in the space region of the bi-soliton having width $\sim 70.0$; meanwhile, the impact of the rest part of the computational domain, where some small wavefield oscillations can also be seen, contributes $\sim 10^{-3} I$, and thus can be neglected. The latter means that the bi-soliton exists on its own and does not participate in resonances with continuous waves typical for some nonlinear wave systems \cite{boyd2012weakly,yang1999embedded,khusnutdinova2018soliton}.

We find that the major dynamics of the DZE bi-solitons can be described by the exact two-soliton solution of the NLSE with dynamically changing eigenvalues and norming constants as
\begin{equation}
\label{2ss_model}
q_{\text{(DZE)}}(x,t_j) = q^{\text{S}}_{2}(x,t_j;\{\lambda_{n}(t_j),\rho_n(t_j)\}).
\end{equation}
Here $\{\lambda_{n}(t_j),\rho_n(t_j)\}$, $n=1,2$ are the set of the computed time series of the discrete IST spectrum. The evolution of model (\ref{2ss_model}) and its comparison with recorded bi-soliton wavefield are shown in Figure~\ref{fig:fig2}.
In contrast, an arbitrary choice of soliton eigenvalues leads to a formation of unstable trajectories that we illustrate in Supplemental Material \cite{SuppM}.

We compute the soliton eigenvalue changes at each time step as $\Delta\lambda(t_j) = \lambda(t_{j+1})-\lambda(t_{j-1})$,
and compare the obtained function $\Delta\lambda(t_j) = \Delta \xi(t_j) + i \Delta\chi(t_j)$ with predictions of the perturbation theory. We find that Eq.~(\ref{dzeta}) provides an excellent description for both solitons, see Figure~\ref{fig:fig3}. To evaluate the integral in Eq.~(\ref{dzeta}) we use numerically computed $\text{RHS}_{\text{(DZE)}}(t_j)$ obtained by substitution $q_{\text{(DZE)}}(x_i,t_j)$ into Eq.~(\ref{RHS_forms}) and the corresponding numerically computed wave functions $\Phi(x_i,t_j)$; see details in Supplemental Material \cite{SuppM}. Our comparison shows that the DZE bi-soliton complex exists in the regime of nearly integrable dynamics.

Finally, we perform the IST analysis of the long-living RVE bi-soliton and obtain qualitatively similar results as for the DZE case. Figure~\ref{fig:fig4}(a) shows the surface elevation in physical units corresponding to the minimum and maximum of the RVE bi-soliton intensity. Applying transformation (\ref{q_transformation}) we obtain the bi-soliton envelope $q_{\text{(RVE)}}(x,t_j)$ oscillating with $T\approx 30.0$ and compute the soliton eigenvalue trajectories, see Figure~\ref{fig:fig4}(b). The trajectories are slightly perturbed in comparison to the DZE case and experience a minor drift on a scale of hundreds of bi-soliton oscillation periods. The rest of the IST analysis repeats our results for the DZE and is presented in Supplemental Material \cite{SuppM}. Note that in the case of the RVE bi-solitons, the IST perturbation theory works only quantitatively, which is expected for the fully nonlinear model due to the presence of the complicated structure of its RHS.

\begin{figure}[!t]\centering
	\includegraphics[width=1.0\linewidth]{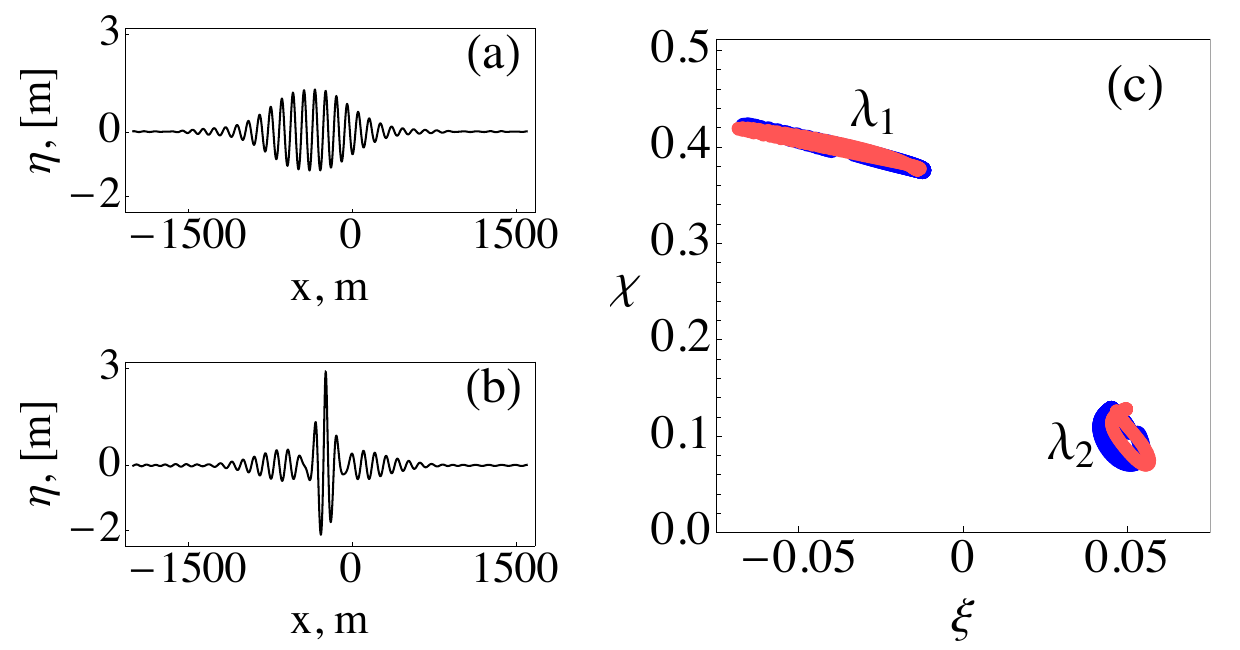}
	\caption{\small 
    Nonlinear behavior of long-living bi-soliton in the RVE. (a,b) Surface profiles $\eta(x)$ at minimum and maximum amplitude respectively. (c) Stable trajectories of soliton eigenvalues. Blue and red lines correspond to two complete cycles of the bi-soliton oscillations separated by $200$ characteristic oscillation periods $T \approx 30.0$.
    }
	\label{fig:fig4}
\end{figure}

The performed IST analysis of the long-living bi-solitons in the deep water models shows that these oscillatory complexes exist near the NLSE integrable regime and can be described within the IST perturbation theory. In general, the governing DZE and RVE equations are far from being integrable; however, for the bi-solitons all the RHS terms of the DZE or RVE are small throughout the whole oscillation period $T$. The numerically computed time series of the IST spectrum allows us to accurately reveal the recursive dynamics of the bi-solitons preserving at the scale of hundreds of $T$. We also show that the bi-soliton complex is stabilized by a nonzero velocity difference between solitons and minor radiation; both of them gradually increase up to the high-amplitude wavefield configuration so that the two discrete components and continuous part of the scattering data are in periodic energy and momentum exchange.

In contrast to the approximately solvable models with weakly interacting solitons, see \cite{karpman1981perturbational,gorshkov1981interactions,gorshkov1979existence,gerdjikov1996asymptotic,yang2001interactions,zhu2007universal}, the bi-solitons considered here are fully overlapping and governed by equations of type (\ref{eqNLS}) with such complicated RHS, that cannot be studied analytically with the perturbation framework (\ref{dzeta}). Here, we propose a perspective of using IST theory in such non-solvable cases based on the combination of the perturbation approach, exact multi-soliton solutions, and numerical IST tools. Our approach provides an IST interpretation of the interaction mechanism for the deep water bi-solitons and opens questions for further studies. One of them is identifying a complete set of initial soliton eigenvalues corresponding to long-living recursive bi-soliton dynamics. Another question concerns the connection of the presented approach with general methods of finding periodic solutions to nonlinear PDEs \cite{ambrose2010computation,wilkening2012overdetermined}. Our approach can be generalized to other physical systems, such as optical waveguides described in the leading order by the NLSE \cite{kivshar2003optical,YangBook}, and also applied to analyze experimental data.

\begin{acknowledgments}
The work of AG was funded by the European Union's Horizon 2020 research and innovation program under the Marie Skłodowska-Curie grant agreement No. 101033047.
The work of SD and DK on obtaining and studying bi-solitons in the deep fluid models was supported by the RSF Grant No. 19-72-30028.
The work of RM on IST perturbation theory analysis was supported by RSF Grant No. 19-79-30075-$\Pi$.
\end{acknowledgments}

%

\end{document}